\magnification=1200
\baselineskip=20pt

\vfill\eject
\def\gm{\gamma}
\def\sqs{\sqrt {s}}
\def\lam{\Lambda}
\def\dmu{D_{\mu}}
\def\dmuphi{D^{\mu}\phi}
\def\delmu{\partial_{\mu}}
\def\delmuh{\partial^{\mu}H}
\def\gf{\gamma_5}
\def\sigmunu{\sigma^{\mu\nu}}
\def\alr{A_{LR}}
\def\gmmuu{\gamma^{\mu}}
\def\lmo{\lambda_1}
\def\lmt{\lambda_2}
\def\lm{\lambda}
\def\ve{\bar {v}(p_2, \lambda_2)}
\def\yu{ u(p_1, \lambda_1)}
\def\eps{\epsilon^*}

\centerline{\bf Effect of Tev scale new physics on the cross-section}
\centerline{\bf for the process $e^+e^-\rightarrow H\gm $}

\vskip .4truein
\centerline{\bf Uma Mahanta}
\centerline{\bf Mehta Research Institute}
\centerline{\bf Chhatnag Road, Jhusi}
\centerline{\bf Allahabad-221506, India}

\vskip 1truein
\centerline{\bf Abstract}

Tev scale new heavy physics could significantly
 affect the cross-section for the rare process
$e^+e^-\rightarrow H\gm $ through non-renormalizable
 operators involving the light SM fields.
In this article we show that for $\sqs =500$ Gev and $\lam =1$ Tev some 
of these operators, which are only weakly constrained by the latest
 LEP and SLD data,
could produce an observable number of events for $m_H$ lying in the range
200-500 Gev. Whereas for $\lam =5$ Tev although no event with
$m_H\le 250$ Gev is likely to be seen, for moderately heavy higgs bosons
 with  $m_H$ lying in the range (400-500) Gev the production rate will 
be large enough to be detectable.

\vfill\eject

New heavy physics at the Tev scale could affect physical processes at
lower energies through non-renormalizable operators invariant under the
$SU(3)_c\times SU(2)_l\times U(1)_y $ gauge symmetry and relevant
global symmetries. The operators should involve the light SM fields
only and could be expressed as a systematic power series expansion
in ${1\over \lam}$. The search for higgs bosons at $e^+e^-$ collider
usually concentrates on the $e^+e^-\rightarrow ZH$ channel [1].
However the process  $e^+e^-\rightarrow \gm H$ channel is
also important since it allows one to extend the range of higgs boson
mass at a given center of mass energy. Further since 
 the cross-section for the process $e^+e^-\rightarrow
H\gm $ is very small in the SM [2,3], the 
observation of any significant number of events at LEP2 or NLC500
would clearly signal the effect of new physics. The lowest dimensional
(d=6) flavor diagonal operators involving leptons, scalar and gauge
fields that can affect the process $e^+e^-\rightarrow H\gm $ at tree level
are [4]

$$\eqalignno{O_1&=(\bar{l}\dmu e_R)\dmuphi +h.c.\cr
&={1\over \sqrt{2}}[\bar{e}_L(\delmu e_R)+(\delmu \bar{e}_R)e_L]\delmuh
-{\imath e\over \sqrt{2}}(\bar{e}\gf e)A_{\mu}\delmuh\cr
&+{\imath g\over 2\sqrt {2}c_w}[(\delmu \bar{e}_R)e_L-\bar{e}_L(\delmu e_R)]
Z^{\mu}(v+H)+{\imath es_w\over \sqrt{2} c_w}(\bar{e}\gf e)Z_{\mu}\delmuh . 
&(1)\cr}$$

$$\eqalignno{O_2&=(\dmu \bar {l})e_R\dmuphi +h.c.\cr
&={1\over \sqrt{2}}[\bar{e}_R(\delmu e_L)+(\delmu \bar{e}_L)e_R]\delmuh
+{\imath e\over \sqrt{2}}(\bar{e}\gf e)A_{\mu}\delmuh\cr
&+{\imath g\over 2\sqrt {2}c_w}[ \bar{e}_R (\delmu e_L)-(\delmu\bar{e}_L) e_R]
Z^{\mu}(v+H)+{\imath e(1-2s^2_w)\over 2\sqrt{2}s_w c_w}
(\bar{e}\gf e)Z_{\mu}\delmuh . &(2)\cr}$$

$$\eqalignno{O_3&=(\bar{l}\sigmunu\tau_a e_R)\phi W^a_{\mu\nu}+h.c.\cr
&=-{1\over \sqrt {2}}(\bar {e}\sigmunu e)(c_w Z_{\mu\nu}+s_w F_{\mu\nu})
(v+H).&(3)\cr}$$
and
$$\eqalignno{O_4&=(\bar{l}\sigmunu e_R)\phi B_{\mu\nu}+h.c.\cr
&={1\over \sqrt {2}}(\bar {e}\sigmunu e)(c_w F_{\mu\nu}-s_w Z_{\mu\nu})
(v+H).&(4)\cr}$$

In the above we have expressed the fields in unitary gauge and have
written only the operators that contain no gauge field or at most
one neutral gauge field. The effective low energy Lagrangian can be written
 as $L_{eff}=L_{sm}+L_{\lam }$ where $L_{\lam }=\sum_{i=1}^4 C_iO_i$.
$C_i^{-{1\over 2}}$ is the characteristic scale associated with the
d=6 operator
$O_i$. On transforming the lepton fields from the gauge basis to the
mass basis, $O_3$ and $O_4$ give rise to d=5 FCNC vertices involving the
 Z boson and photon. In particular $C_3$ and $C_4$ must satisfy the strong
 experimental bound [5] on the branching fraction for the FCNC
 process $\mu\rightarrow e\gm $ which implies that

$${\Gamma_{\mu\rightarrow e\gm }\over \Gamma_{\mu\rightarrow e\bar{\nu}_e
\nu_{\mu}}}\approx {6\pi^2 v^6\over \lam ^4 m^2_{\mu}}(c_w-s_w)^2
\sin^2\theta_{12}\le 5\times 10^{-11}.\eqno(5)$$

If we assume that $\sin\theta_{12}\approx .2$, the scale $\lam\approx
C_3^{-{1\over 2}}\approx C_4^{-{1\over 2}}$ associated with $O_3$
and $O_4$ must be greater than 3500 Tev. Such a gigiantic
 scale for $O_3$ and $O_4$ can be
 avoided by assuming a flavor symmetry that forbids FCNC vertices 
when the leptons are transformed
from the gauge basis to the mass basis. However the flavor diagonal terms 
of $O_3$ and $O_4$ must still satisfy the strong constraint arising
from the experimental value [6] of ${g_e-2\over 2}$ which implies that

$$({\delta g_e\over 2})_{expt}-({\delta g_e\over 2})_{sm}=
({\delta g_e\over 2})_{new}\approx {2\sqrt {2} m_e v(c_w-s_w)\over \lam^2 e}
\le .27\times 10^{-9}.\eqno(6)$$

Hence the scale associated with the flavor diagonal terms of $O_3$
and $O_4$ must be greater than 40 Tev. In the following we shall
therefore ignore the contribution of $O_3$ and $O_4$ to the cross-section
for the process $e^+e^-\rightarrow H\gm $.
 In contrast $O_1$ and $O_2$
do not contain any $ee\gm $ vertex but they do contain eeZ vertex.
The scale associated with these operators can be best constrained by
Z pole precision measurements. $O_1$ for example affects $Z$ pole physics 
through the Lagrangian
 $$L_{\lam }={-\imath gv\over 2\sqrt {2}c_w \lam^2}
[\bar{e}_L(\delmu e_R)-(\delmu \bar{e}_R)e_L]Z^{\mu}.\eqno(7)$$
 The new physics 
contribution to $A_{LR}$ can in general be written as

$${A_{LR}-(A_{LR})_{sm}\over (A_{LR})_{sm}}\approx {\delta\sigma_L-\delta
\sigma_R \over (\sigma_L-\sigma_R)_{sm}}- {\delta\sigma_L+\delta
\sigma_R \over (\sigma_L+\sigma_R)_{sm}}.\eqno(8)$$

Evaluating the contribution of $O_1$ to $\sigma_L$ and $\sigma_R$ on
Z pole we find that $ {\delta\sigma_L-\delta
\sigma_R \over (\sigma_L-\sigma_R)_{sm}}\approx {1\over 2}{({g\over 4c_w})^2
{v^2s\over \lam^4}\over (g^e_L)^2+ (g^e_R)^2}$ and
$ {\delta\sigma_L+\delta
\sigma_R \over (\sigma_L+\sigma_R)_{sm}}\approx {({g\over 4c_w})^2
{v^2s\over \lam^4}\over (g^e_L)^2+ (g^e_R)^2}$. The contribution of
$O_1$ to $A_{LR}$ is therefore given by ${\delta \alr \over (\alr )_{sm}}
\approx - {1\over 2}{({g\over 4c_w})^2
{v^2s\over \lam^4}\over (g^e_L)^2+ (g^e_R)^2}$.
 The present SLD precision [7] for measuring
$\alr $ is about 5\%. Hence even for $\lam $ as low as 400 Gev, the change
$\delta \alr $ is far too small (about .78\%) to be detected at SLD. 
Similar conclusion can be reached by analysing the effect of $O_1$
on other Z pole observables for example $A^e_{fb}$.
The scale $\lam $ associated with $O_1$ and $O_2$ is therefore rather weakly
constrained by existing experimenta data. Further since $O_1$ and
$O_2$  are similar in structure in the following we shall consider
only the effect of $O_1$ on the process $e^+e^-\rightarrow H\gm $.

The relevant effective Lagrangian that determines the cross-section for the
process  $e^+e^-\rightarrow H\gm $ is

$$\eqalignno{L_{eff}&=L_{sm}+L_{\lam }\cr
&=\bar {e}(\imath \delmu \gmmuu +eA_{\mu}\gmmuu )+{1\over \sqrt {2}\lam^2}
[\bar {e}_L(\delmu e_R)+(\delmu \bar {e}_R)e_L]\delmuh\cr
&+{\imath e\over \sqrt {2} \lam^2}\bar {e}\gf e A_{\mu}\delmuh .&(9)\cr}$$

The first thing to note is that both $L_{sm}$ and $L_{\lam}$ are
separately invariant w.r.t. $U(1)_Q$ gauge transformations. Second
the eeH vertex of  $L_{\lam}$ is of order ${1\over \lam^2}$ but the 
$ee\gm H$ vertex is of order ${e\over \lam^2}$. The invariant matrix
element for the process $e^+e^-\rightarrow \gm H$ can be written as
$ M=M_1+M_2+M_3 $ where

$$M_1=-{e\over 4\sqrt {2}\lam^2 }\ve [m^2_H+t\gf ]{2(p_1.\eps_\lm )
-(q.\gm )(\eps_\lm .\gm )\over p_1.q}\yu .\eqno(10a)$$

$$M_2 = -{e\over 4\sqrt {2}\lam^2 }\ve{(\eps_\lm .\gm)(q.\gm )-
2(\eps_\lm .p_2)\over p_2.q} [m^2_H-u\gf ]\yu .\eqno(10b)$$
and
$$M_3=-{e\over \sqrt {2}\lam^2}\ve \gf\yu \eps_\lm .p_3 .\eqno(10c)$$

Here q and $\lm $ are the momentum and helicity of the outgoing photon
and $p_3$ is the momentum of the outgoing higgs boson. 
$t=(p_1-q)^2=(p_2-p_3)^2$
and $u=(p_2-q)^2=(p_1-p_3)^2$. $M_1$ and $M_2$ arise from the product
of $L_{sm}$ and $L_{\lam}$. Whereas $M_3$ arises from $L_{\lam}$ only.
Note that the expression for M vanishes if we replace $\epsilon_{\mu}$ by
$q_{\mu}\theta$, which guarantees its invariance w.r.t. $U(1)_Q$ gauge
 transformations. The square of the invariant matrix element averaged over the
 polarizations of the incoming $e^-$ and $e^+$  
and summed over the polarization of the outgoing $\gm$ is given by
${1\over 4}\sum_{\lmo ,\lmt , \lm}\vert M\vert ^2 ={e^2m_H^4\over 8\lam^4}
{m_H^4+s^2\over ut}$. The total cross-section integrated over all directions
is infinite. This infinity
 can be avoided by imposing a rapidity cut (for example  $y=
\ln\cot{\theta_{\min}\over 2}=2.6)$ on the outgoing $\gm $. The total 
unpolarized cross-section for the process $e^+e^-\rightarrow H\gm $ then
becomes
$$\sigma ={\alpha m_H^4\over 8 \lam^4}{1\over s}{m_H^4+s^2\over s(s-m_H^2)}
y .\eqno(11)$$   
 
Note first that the cross-section depends very strongly on $m_H$ and $\lam $.
The rapid rise of the new physics contribution with increasing $m_H$
 could be used to
distinguish it from the SM contribution which falls off rapidly with $m_H$.
Second as $m_H^2\rightarrow s$ the cross-section becomes singular as
${1\over (s-m_H^2)}$. The reason is that as $m_H^2\rightarrow s$ the
energy of the outgoing
photon approaches zero and
the electron propagator becomes on shell giving rise to a factor
${1\over (s-m_H^2)^2}$ in $\vert M\vert ^2$.
 The phase space suppression factor
of $s-m_H^2 $ is not sufficient to tame this divergence. However
the process $e^+e^-\rightarrow H\gm $ with a zero energy photon cannot be
distinguished from $e^+e^-\rightarrow H$. In fact the IR singularity
 of the former is precisely cancelled by the IR singularity of the
O($\alpha$) correction to $e^+e^-\rightarrow H$. 
In order to overcome this infrared
singularity one can impose a lower bound on the outgoing photon energy.
Further for a given $\lam $ and $\sqrt{s}$ an upper bound on $m_H$ can
be determined from the unitary bound on $\vert M\vert ^2$
which requires that ${2\pi e^2 y m_H^4\over \lam ^4}{m_H^4+s^2\over 
(s-m_H^2)^2}<1$.
 For $\lam =1$
Tev and $\sqrt{s}= 500$ Gev the unitary bound on $m_H$ is roughly 400 Gev
and it increases with increasing cut off.

LEP2 is expected to operate at $\sqrt{s}\approx 180 $Gev and  with an
intergrated luminosity of (.5-1)fb$^{-1}$/yr. For $m_H=150 $ Gev, $\sqs $=200
Gev and 
$\lam =1$ Tev, $\sigma$ turns out to be 36.7 fb and therefore a few tens of
events are likely to be seen at LEP2. However if the cut off is raised
to 5 Tev $\sigma $ drops to .06 fb and then no events are expected.
In contrast in the context of SM, for $m_H=$ 150 Gev the unpolarized 
cross-section is expected to be .04 fb [2] at LEP2 total energy.
NLC 500 is expected to achieve a luminosity of around 50 fb$^{-1}$/yr
in its final stage. We find that for a higgs mass of 250 Gev, $\sigma$ =22
fb and .04 fb corresponding to cut off scales of 1 Tev and 5 Tev
respectively. However for a moderately heavy higgs with $m_H $=400 Gev,
$\sigma$ is given by 400 fb and .64 fb corresponding
 to new physics scales of
1 Tev and 5 Tev respectively. At $\sqrt {s}$=500 Gev the corresponding
SM contributions [3] are .03 fb (for $m_H$=250 Gev)  and .003 fb
 (for $m_H$=400 Gev). We can therefore conclude that if the new physics 
scale is as low as 1 Tev, then NLC500 will produce a significant number of 
events with $m_H$ lying in the range 200-400 Gev. However if $\lam =$ 5 Tev
then although no events with light higgs bosons are expected, a few tens of 
events with moderately heavy higgs bosons ($400\  Gev < m_H <  500\  Gev$)
 are likely to be seen.

In conclusion in this article we have considered the effect of a d=6
operator on the cross-section for the process 
$e^+e^-\rightarrow H\gm $. We have shown that the scale associated with
the operator is rather weakly constrained by the existing experimental data.
In fact $\lam $ can be as low as 400 Gev without causing detectable effects on 
 Z pole precision measurements. For $\lam $=1 Tev the new physics 
contribution to the cross-section is large enough to be seen at NLC500
for $m_H$ lying in the range 200-400 Gev. On the other hand for $\lam $=5
Tev although no events with light higgs bosons are expected, an observable
number of events with moderately heavy higgs bosons ($400\  Gev< m_H < 500\ 
 Gev$) are likely to be produced.

\vfill\eject 
\centerline{\bf References}

\item{1.} J. Gunion, H. Haber, G. Kane and S. Dawson, ``The Higgs
Hunter's Guide'' ( Addision-Wesley, Reading, 1990).

\item{2.} A. Barroso, J. Pulido and J. C. Romao, Nucl. Phys.
 B 267, 509 (1985); ibid. B 272, 693 (1986); A. Abbasabadi, D. B. Chao,
D. A. Dicus and W. A. Repko, Phys. Rev. D 52, 3919 (1995).

\item{3.} A. Djouadi, V. Driesen, W. Hollik and J. Rosiek, Nucl. Phys. B
491, 68 (1997).

\item{4.} W. Buchmuller and D. Wyler, Nucl. Phys. B 268, 621 (1986).

\item{5.} Review of Particle properties, Phys. Rev. D 54, 21 (1996).

\item{6.} T. Kinoshita, Phys. Rev. Lett. 47, 1573 (1981); Review of Particle
 properties, Phys. Rev. D 54, 21 (1996).

\item{7.} Reiview of Paricle Properties, Phys. Rev. D 54, 222 (1996).

\end